\documentstyle[twocolumn,aps]{revtex}

\input psfig.sty

\def\ltsim{\vbox {\hbox{\lower 0.6\baselineskip \hbox{$<$}} \break
		 \hbox{\lower 0.1\baselineskip \hbox{$\sim$}} } }
\def\gtsim{\vbox {\hbox{\lower 0.6\baselineskip \hbox{$>$}} \break
                 \hbox{\lower 0.1\baselineskip \hbox{$\sim$}} } }
\def\k{{\bf k}}

\def\r{{\bf r}}

\def\vs{{\bf v}_s}
\begin{document}
\draft

\twocolumn[\hsize\textwidth\columnwidth\hsize\csname %
@twocolumnfalse\endcsname

\title{Quasiparticles in the Vortex State of Dirty $d$-Wave Superconductors}

\author{
C. K\"ubert and P.J. Hirschfeld
}

\address{
Department of Physics, University of Florida, Gainesville, FL 32611, 
USA.\\
}

\maketitle
\begin{abstract}
We consider the problem of the vortex contribution to thermal properties
of dirty $d$-wave superconductors.  In the clean limit, Volovik has argued
that the main contribution to the density of states in a $d$-wave
superconductor arises from extended quasiparticle states
which may be treated semiclassically, giving rise to a
specific heat contribution $\delta C(H)\ltsim H^{1/2}$ .  We show that the
extended states continue to dominate the dirty limit, but lead to a $H
\log H$ behavior at the lowest fields, $H_{c1}\ll H\ll H_{c2}$.
This crossover may explain recent discrepancies in specicific heat
measurements at low temperatures and fields in the cuprate superconductors.
We further discuss the field dependence
of transport properties within the same model.
\end{abstract}

]
{\it Introduction.} 
As pointed out by Volovik,\cite{Volovik1} $d$-wave superconductors are
fundamentally different from their $s$-wave counterparts in the vortex
state,  in that at low temperatures and fields the extended quasiparticle
states around the vortex play a much more important role.   In the case
of the single-particle density of states, he showed in fact that the
contribution
from these states over the entire vortex unit cell exceeds that of the
bound states.  It is therefore to be expected that the quasiparticles
will also play an important role in transport properties of a
$d$-wave system in the presence of an applied magnetic field.  To
study low-$T$ transport, one needs a way of including scattering
by defects at a simple level.  We recently studied the specific heat
$\delta C (H)$ of $d_{x^2-y^2}$ superconductors in a field using
the common $t$-matrix approximation for strong impurity
scattering,\cite{KH} proposing a possible explanation for
deviations of some measurements from the $\delta C\sim \sqrt{H}$
behavior predicted for a clean $d$-wave superconductor by Volovik.
We also showed how other consequences of the clean Dirac spectrum,
including the $1/\omega$ divergence in the density of states
$\delta N(\omega; H)$\cite{KopninVolovik1} and the predicted
scaling\cite{SimonLee}
of the specific heat $\delta C/(TH^{1/2}) =F_C(H^{1/2}/T)$, are
modified by disorder.
    \vskip .2cm
In this paper we review the results on thermodynamic properties
and present a preliminary calculation of the low-field microwave conductivity.
 We neglect other possible contributions
to the conductivity arising from disordered vortex scattering, and
assume that the Abrikosov lattice is pinned, ordered, and sufficiently
dilute to not alter the quasiparticle bands significantly from their
1-vortex form.
We point out that confirmation of the magnetic field and disorder
dependence would not only provide evidence for the primacy of
quasiparticle transport in $d$-wave superconductors, but also
yield information on the phase shifts of the defects and on
quasiparticle relaxation times.
     \vskip .2cm
{\it Semiclassical treatment of extended states.}
In the semiclassical approximation the Doppler effect associated with 
supercurrents around a vortex lead to a shift of the quasiparticle energy
$\omega\rightarrow\omega-\vs\cdot\k$, where $\vs= (\hbar/2mr){\hat \theta}$
is the superfluid velocity. 
The impurity-averaged electron propagator 
in particle-hole space is  given by 
\begin{eqnarray}
g(\k,\omega; \vs) = \frac
{  ( \tilde \omega-\vs\cdot\k) \tau_0 +  \Delta_{\k}\tau_1
                         +  \xi_{\k} \tau_3 }
{(\tilde \omega-\vs\cdot\k)^2 -  \Delta_\k^2 -\xi_\k^2} \; ,
\end{eqnarray} 
where the $\tau_i$ are  Pauli matrices.
Due to symmetry of the $d_{x^2-y^2}$ order parameter 
$\Delta_\k=\Delta_0\cos 2\phi$
and the assumed particle-hole symmetry of the normal state only the frequency
is renormalized $\tilde \omega = \omega-\Sigma_0 (\omega)$. 
In the unitarity limit the self-energy is given by
$\Sigma_0 = \tau_0 \Gamma/G_0$ 
where $\Gamma=n_i/\pi N_0$ is an impurity scattering rate depending on
the concentration $n_i$ of point potential scatterers and 
the density of states at the Fermi level, $N_0$.
For a simple $d_{x^2-y^2}$ superconductor the averaged integrated 
Green's function reads
$G_0(\omega,\vs)=-i(2/\pi) {\bf K}(\Delta_0/(\tilde\omega-\vs\cdot\k))$,
where ${\bf K}$ is the complete elliptic integral of the first kind.
\begin{figure}[h]
\leavevmode\centering\psfig{file=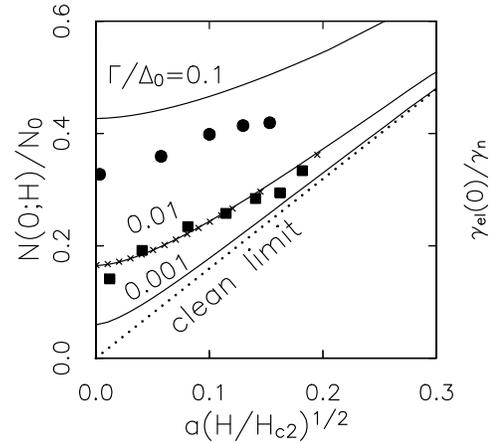,width=2.6 in}
\caption{$N(0;H)/N_0$ for
$\Gamma/\Delta_0=0.1$, 0.01, and 0.001 in unitarity limit (solid lines)
and the clean limit (dotted line).
Data from Fisher et al.\protect\cite{Fisheretal} (circles);
Moler et al.\protect\cite{Moleretal} (untwinned sample, squares),
assuming $H_{c2}/a^2$=300T,
$\gamma_n=15$ mJ-mol-${\rm K}^2$. }
\end{figure}
{\it Density of states at zero energy.}
To find $N(0;H)$, we average
the propagator over a vortex unit cell, $N(0;H)/N_0\equiv \langle- {\rm Im}
G_0(\omega;\vs)\rangle_H$, where for any $f(\vs)$ we define
$\langle  f(\vs) \rangle_H\equiv A^{-1}(H)\int_{\rm cell} d^2r\;  f(\vs )$.
The magnetic field dependence enters only through the inter-vortex spacing
$R = \xi_0 (\pi/2)^{1/2} a^{-1} (H_{c2}/H)^{1/2}$.
In the clean limit we reproduce the result of Volovik
$N(0;H)/N_0\simeq \sqrt{8/\pi}a(H/H_{c2})^{1/2}$.
In the dirty limit, where the zero-energy quasiparticle scattering rate 
$\gamma_0$ is much larger than the average quasiparticle energy shift
$E_H\equiv a(H/H_{c2})^{1/2}\Delta_0$, we find
\begin{equation}
\frac{\delta N(0,H)}{N_0}\simeq {\Delta_0\over 8\gamma_0}
a^2\left ({H\over H_{c2} } \right) \log
\left[ {\pi\over 2a^{2}}\left ({H_{c2}\over H}  \right)  \right]
\end{equation}
Numerical results for the density of states $N(0,H)$ together with 
experimental data of Moler et al.\cite{Moleretal} and 
Fisher et al.\cite{Fisheretal} are given in Fig.~1.
The data Moler et al.~are consistent with a slightly 
dirty d-wave superconductor, while the Fisher data cannot be well fitted 
to the clean case, suggesting that their sample contains roughly an order of 
magnitude more defects.
     \vskip 0.2 cm
{\it Density of states at finite frequency.} 
The field dependent part of the density of states for 
$v_sk_F\ll \omega\ll \Delta_0$ is given by 
$\langle -{\rm Im}[G_0(\omega;\vs)-G_0(\omega)] \rangle_H$\cite{KopninVolovik1}
which for $ \gamma_0,E_H < \omega$ yields 
\begin{eqnarray}
&&\frac{\delta N(\omega;H)}{N_0}\simeq \langle\left[
\left|{\vs\cdot\k_n-\omega\over \Delta_0}\right| - \left|{\omega\over
\Delta_0}\right|\right]
\rangle_H
\simeq {E_H\over \sqrt{2\pi\Delta_0^2}}F(x)\\&&F(x)\equiv{1\over x}\left\{
\begin{array}{ll}
\pi/ 2 & x >1\\
3x\sqrt{1-x^2}+(1+2x^2)\sin^{-1}x-\pi x^2 ~&  x<1
\end{array}\right.~~~\nonumber
\end{eqnarray}
where $x=\sqrt{2/\pi}(\omega/ E_H)$ and $\k_n$ the nodal direction.

In Fig.~2 we plot the density of states 
versus frequency. 
In the intermediate range we recover the predicted $1/\omega$
divergence $\delta N(\omega;H)/N_0\simeq a^2\pi\Delta_0H/(4\omega
H_{c2})$, which in the clean limit $\gamma_0 < E_H$ 
is cut off as $\delta N(\omega;H)\simeq N(0;H)(1-\pi x/4))$ and
in the dirty case by the impurity scattering scale $\gamma_0 > E_H$.
\begin{figure}[h]
\leavevmode\centering\psfig{file=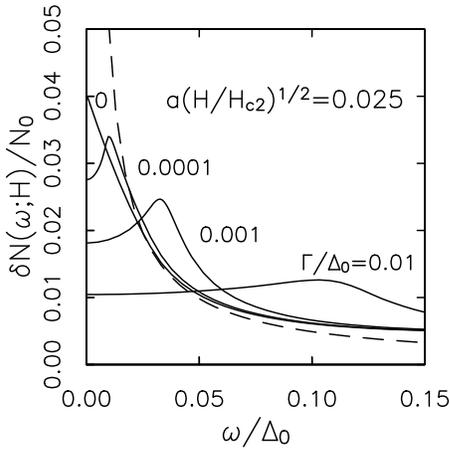,width=2.6 in}
\caption{Density of states at $a(H/H_{c2})^{1/2}=0.025$ vs.
freq. $\omega/\Delta_0$ for $\Gamma/T_c$=0.,0.0001,0.001, 0.01 (solid
lines).  Intermediate-frequency asymptotic result $\delta N \simeq
a^2\pi \Delta_0 H/(4 \omega H_{c2})$ (dashed line).}
\end{figure}

{\it Scaling of specific heat}.
The specific heat at low-temperature is given by
\begin{eqnarray}
C&\simeq&2\int_0^\infty d\omega \left(\omega\over T\right)^2
\left({-\partial f\over \partial \omega} \right)  N(\omega;H)
\\
&\simeq&\left\{ \begin{array}{ll}  N(0;H) {\pi^2\over 3} T
~~~ &T \ll {\rm max}[\gamma_0,E_H]\ll \Delta_0 \\
N_0 \left(9\zeta(3) T^2\over \Delta_0 \right)~~~~ & \gamma_0,E_H\ll T \ll \Delta
_0
\end{array}\right.\nonumber
\end{eqnarray}
A $T^2$ term  characteristic of the pure $d$-wave system in zero field is
present whenever both the impurity and magnetic field scales are smaller
than the temperature.
Substituting Eq.~(3) into (4) leads to the scaling function of the 
specific heat  
$\delta C(H)/[\gamma_nTa(H/H_{c2})^{1/2}]=F_C[Y]$,
where $Y=a(H/H_{c2})^{1/2}T_c/T$.  The scaling function $F_C$ 
varies as $F_C\simeq 3\log 2 \Delta_0 Y/(4\pi T_c)$ for $Y\ll 1$ and as
$F_C\simeq \sqrt{2/\pi}$ for $Y\gg 1$. 
Scaling is expected for a given data set
provided $H,T$ are such that $E_H$ and $T$
are both larger than the impurity scale $\gamma_0$.
A full numerical evaluation of (4) plotted in Fig.~3 shows that 
for the clean case (open symbols), scaling is
obtained over the full range of $Y$, whereas for the dirty
system (filled symbols) scaling has broken down completely.
\begin{figure}[h]
\leavevmode\centering\psfig{file=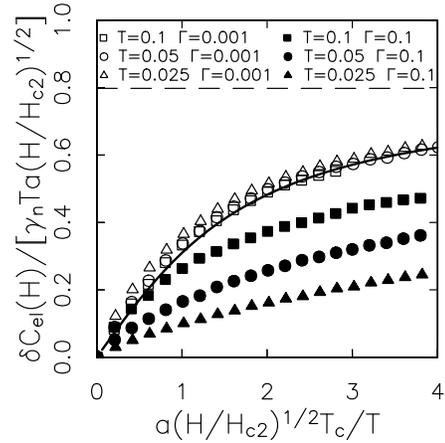,width=2.6 in}
\caption{Normalized vortex contribution to specific heat,
$\delta C_{el}(H)/[\gamma_{el}Ta(H/H_{c2})^{1/2}]$ vs.
$Y\equiv a(H/H_{c2})^{1/2}T_c/T$ for fixed temperatures $T$ and
scattering rates $\Gamma$ as shown; unit of energy $T_{c0}$.
Asymptotic large-$Y$ limit $\sqrt{2/\pi}$ (dashed line).
}
\end{figure}

{\it Microwave conductivity.} 
The microwave conductivity is given by\cite{pjh}
\begin{eqnarray}
\sigma_{ij} (\Omega) = -\frac{ne^2}{m}
\int_{-\infty}^{+\infty} d\omega \:
\left[
\frac{\beta}{4} {\rm sech}^2 (\frac{\beta \omega}{2})
\cdot S_{ij} (\omega,\Omega)
\right] \; ,
\end{eqnarray}
where
\begin{eqnarray}
S_{ij} (\omega,\Omega) &=& {\rm Im} \int \frac{d\phi}{2\pi} \: \hat k_i \hat k_j
\cdot
\left[
  \frac{\tilde \omega_{+}'}{\tilde \omega_{+} - \tilde \omega_{+}'}
  \left(\frac{1}{\xi_{0+}'} - \frac{1}{\xi_{0+}}\right)
\right.
\nonumber \\
&+&
\left.
  \frac{\tilde \omega_{-}'}{\tilde \omega_{+} - \tilde \omega_{-}'}
  \left(\frac{1}{\xi_{0+}} + \frac{1}{\xi_{0-}'}\right)
\right] \; .
\end{eqnarray}
Here we have defined
$\tilde\omega_{\pm}= \bar \omega -\Sigma_0(\omega\pm i0^+)$,
$\xi_{0\pm}=
\pm\mathop{\rm sgn}(\bar \omega) 
({\tilde\omega}_{\pm}^2-\Delta_k^2)^{1/2}$,
as well as analogous primed quantities
$\tilde\omega_{\pm}^\prime=\tilde\omega_{\pm}(\bar \omega-\Omega)$
and $\xi_{0\pm}^\prime=\xi_{0\pm}(\bar \omega-\Omega)$, 
with $\bar \omega = \omega - \vs \cdot \k_n$.
Taking the limit $\Omega \rightarrow 0$, $T \rightarrow 0$ and 
performing the $\phi$-integration we find for the contribution to the 
conductivity from currents at a position $\r$ relative to 
the vortex center
\begin{eqnarray}
S_{ii} (\omega,\r)
&=& \frac{1}{4} \sum_{nodes} \frac{1}{\pi\Delta_0} 
{\rm Re} 
\left\{
k \left[ \frac{\bar \omega }{\Sigma_0} {\bf K} (k) - 
         {\bf E} (k) \right]
\right\} \; ,
\end{eqnarray}
where ${\bf K} (k)$ and ${\bf E} (k)$ are the complete elliptic integrals
of the first and second kind, respectively, with the argument 
$k=\Delta_0/(\Delta_0^2 - (\Sigma_0 - \bar \omega )^2)^{1/2}$.
Note the space dependence arises exclusively through the Doppler
shift $\vs \cdot \k_n = (\hbar v_F/2r) \cos \theta$.
At low temperature and for $H_{c1} \ll H \ll H_{c2}$ the leading order 
is given by
\begin{eqnarray}
\sigma(\vs) &=& \frac{\sigma_{00}}{4} \sum_{nodes} 
\int_{-\infty}^{+\infty} d\omega \:
\frac{\beta}{4} {\rm sech}^2(\frac{\beta \omega}{2})
\nonumber \\
&\cdot&
\left\{
1 + \left[
      \left(\frac{\Sigma_0'' \bar \omega}{\Delta_0^2}\right)
    + \left( \frac{\bar \omega}{\Sigma_0''} \right)
    \right] \:
\arctan \left(\frac{\bar \omega}{\Sigma_0''}\right)
\right.
\nonumber \\
&+&
\left.
\frac{1}{2} \left( \frac{{\Sigma_0''}^2 - \bar \omega^2}{\Delta_0^2} \right)
\left[
    \ln \left( \frac{4\Delta_0} {\sqrt{{\Sigma_0''}^2 + \bar \omega^2}} \right)
     - \frac{3}{2}
\right]
\right\} \; ,
\end{eqnarray}
where $\sigma_{00} = \frac{ne^2}{m} \: \frac{1}{\pi \Delta_0}$ 
is the universal conductivity\cite{Lee}. 
In the gapless regime the self-energy reduces to
$-\Sigma_0'' \sim  \gamma_0 \gg E_H $ which leads to 
the following leading order contribution of the magnetic field 
to the conductivity 
\begin{eqnarray}
\delta \sigma (\r) = \sigma_{00} \:
\left(\frac{\vs \cdot \k_n}{\gamma_0} \right)^2 \; .
\end{eqnarray}
In the clean limit $ \gamma_0 \ll E_H $ the self-energy 
$\Sigma_0 (\omega = 0, \r) $ is given by
\begin{eqnarray}
\Sigma_0'' (\omega = 0, \r) \sim
- \Gamma \left| \frac{\Delta_0}{\vs \cdot \k_n} \right|
\end{eqnarray}
which leads to
\begin{eqnarray}
\delta \sigma (\r) = \sigma_{00} \:
\frac{\pi}{2} \: \left( \frac{\vs \cdot \k_n}{\Delta_0} \right)^2
\frac{\Delta_0}{\Gamma} \; .
\end{eqnarray}
By averaging the above results over a unit vortex cell we obtain 
the magnetic field contribution to the zero-frequency conductivity for a
d-wave superconductor
\begin{eqnarray}
\frac{\delta\sigma}{\sigma_{00}} =
\left\{
        \begin{array}{ll}
   \frac{\pi}{4} \:
    \left( \frac{\Delta_0}{\gamma_0} \right)^2 \:
    a^2 \left( \frac{H}{H_{c2}} \right)
    \ln \left( \frac{\pi}{2a^2} \: \frac{H_{c2}}{H} \right)
  &  \gamma_0 \gg E_H \\
 \frac{\pi^2}{8} \:
    \frac{\Delta_0}{\Gamma} \: a^2 \left( \frac{H}{H_{c2}} \right)
    \ln \left( \frac{\pi}{2a^2} \: \frac{H_{c2}}{H} \right)
& \gamma_0 \ll E_H
        \end{array}
\right. 
\end{eqnarray}
These results
should be valid provided the quasiparticle mean free path is in fact limited
by impurities and not by fluctuations in the vortex lattice.  The experiments
of Orenstein et al.~\cite{Orenstein} at 150 GHz and a few Tesla are in 
precisely the correct regime to allow one to neglect absorption 
into vortex oscillation modes, and to compare to our $\Omega \rightarrow 0$
result.  Furthermore, we 
expect even in the presence of elastic (pinned) vortex lattice 
disorder that for  small fields of order a few Tesla, vortices will be sufficiently
dilute that scattering at low $T$ will be impurity dominated. 
Orenstein et al.~indeed observe a convex downward curvature in
$\delta \sigma (H)$, as well as a broad maximum at around $3 {\rm T}$.
Further investigation of scattering of quasiparticles by vortices and
of the effect of applied field on  inelastic spin fluctuations is needed 
to extend our results to higher temperatures and fields.
\vskip 2 mm
{\it Note added:} Results for $N(0)$ in the dirty limit similar to ours 
    were obtained independently by Barash et al. \cite{Barash}
\vskip 2 mm
{\it Acknowledgments.}  
Partial support was provided by
NSF-DMR-96--00105 and by the A. v. Humboldt Foundation (CK).

\end{document}